\documentclass[preprint2]{aastex}

\slugcomment{Not to appear in Nonlearned J., 45.}

\shorttitle{Analysis of a global Moreton wave observed on October
28, 2003} \shortauthors{Muhr et al.}

\begin{document}


\title{Analysis of a global Moreton wave observed on October 28, 2003}


\author{N. Muhr}
\affil{Institute of Physics, University of Graz, Universit\"atsplatz
5 , A--8010 Graz, Austria\\
Hvar Observatory, Faculty of Geodesy, University of Zagreb,
Ka\v{c}i\'{c}eva 26, HR--10000 Zagreb, Croatia}
\email{muhrn@edu.uni-graz.at}

\and

\author{B. Vr\v snak}
\affil{Hvar Observatory, Faculty of Geodesy, University of Zagreb,
Ka\v{c}i\'{c}eva 26, HR--10000 Zagreb, Croatia}
\email{bvrsnak@gmail.com}

\and

\author{M. Temmer and A.~M. Veronig}
\affil{Institute of Physics, University of Graz, Universit\"atsplatz
5 , A--8010 Graz, Austria} \email{mat@igam.uni-graz.at,
asv@igam.uni-graz.at}

\and

\author{J. Magdaleni{\'c}}
\affil{Royal Observatory of Belgium, Ringlaan 3 Avenue Circulaire, B-1180 Brussels, Belgium\\
Hvar Observatory, Faculty of Geodesy, University of Zagreb,
Ka\v{c}i\'{c}eva 26, HR--10000 Zagreb, Croatia}
\email{mjasmina@geof.hr}



\begin{abstract}
    We study the well pronounced Moreton wave that occurred in association with the X17.2 flare/CME event of October 28, 2003.
    This Moreton wave is striking for its global propagation and two separate wave centers, which implies that
    two waves were launched simultaneously. The mean velocity of the Moreton wave, tracked within different
    sectors of propagation direction, lies in the range of $v\approx900-1100$~km~s$^{-1}$ with two sectors showing
    wave deceleration. The perturbation profile analysis of the wave indicates amplitude growth followed by amplitude
    weakening and broadening of the perturbation profile, which is consistent with a disturbance first driven
    and then evolving into a freely propagating wave. The EIT wavefront is found to lie on the same kinematical
    curve as the Moreton wavefronts indicating that both are different signatures of the same physical process.
    Bipolar coronal dimmings are observed on the same opposite East-West edges of the active region as the Moreton wave ignition centers.
    The radio type II source, which is co-spatially located with the first wave
    front,
    indicates that the wave was launched from an extended source region ($\gtrsim 60$~Mm).
    These findings suggest that the Moreton wave is initiated by the CME expanding flanks.
\end{abstract}


\keywords{Shock waves --
             Sun: flares --
             Sun: chromosphere --
             Sun: corona --
             Sun: solar-terrestrial relations}

\section{Introduction}

Large-scale, large-amplitude disturbances propagating in the solar
atmosphere were first recorded by \cite{Moreton:1960b} and
\cite{Moreton:1960} in the chromospheric H${\alpha}$ spectral line;
therefore called ``Moreton waves''. Typical velocities are in the
range $\approx500-1000$~km~s$^{-1}$, and the angular extents are
$\approx90-130\degr$ \citep[e.g.][]{Warmuth:2004,Veronig:2006}.
Since there is no chromospheric wave mode which can propagate at
such high speeds, Moreton waves were interpreted as the intersection
line of an expanding, coronal fast-mode shock wave and the
chromosphere which is compressed and pushed downward by the
increased pressure behind the coronal shockfront
\citep{Uchida:1968}. They occur in association with major flare/CME
events and type II bursts \citep{Smith:1971}, the latter being a
direct signature of coronal shock waves. Typically, the first
wavefront appears at distances of $\ge$100~Mm from the source site
and shows a circular curvature. The fronts are seen in emission in
the center and in the blue wing of the H$\alpha$ spectral line,
whereas in the red wing they appear in absorption. This behavior was
interpreted as a downward motion of the chromospheric plasma with
typical velocities of $\approx 5-10$~km~s$^{-1}$
\citep{Svetska:1976}. Sometimes the trailing segment of the wave
shows the upward relaxation of the material, i.e., the chromosphere
executes a down-up swing \citep{Warmuth:2004b}. In the beginning of
the wave propagation, the leading edge is rather sharp and intense.
As the disturbance propagates, the perturbation becomes more
irregular and diffuse and its profile broadens \citep{Warmuth:2001,
Warmuth:2004}. Thus, the perturbation amplitude decreases and the
wavefronts get fainter until they can no longer be traced at
$\approx300-500$~Mm from the source active region \citep{Smith:1971,
Warmuth:2004}.

A decade ago, large-scale waves were for the first time directly
imaged in the corona by EIT \citep[Extreme-Ultraviolet Imaging
Telescope;][]{Delaboudiniere:1995} aboard the SoHO (Solar and
Heliospheric Observatory) spacecraft, so-called ``EIT waves''
\citep{Thompson:1997}. Similarities in the propagation
characteristics led to the assumption that in at least a fraction of
the events are the EIT waves the coronal counterpart of the
chromospheric Moreton waves \citep[e.g.][]{Thompson:1997,
Warmuth:2001, Vrsnak:2002, Vrsnak:2006, Veronig:2006}. Basic
questions regarding the nature of Moreton and EIT waves are whether
they are caused by the same or by different disturbances, and
whether they are initiated by the associated flare or the CME. For
recent reviews, we refer to \cite{Chen:2005}, \cite{Mann:2005},
\cite{Warmuth:2007} and \cite{Vrsnak:2008}.

Here, we study the fast and globally propagating Moreton wave that
occurred in association with the powerful X17.2/4B flare and fast
CME event from the NOAA AR10486 (S16$\degr$, E08$\degr$) on October
28, 2003. Due to its extreme powerfulness and geo-effectiveness,
diverse aspects of this flare/CME event have been analyzed in a
number of studies \citep[e.g.][]{Gopalswamy:2005, Klassen:2005,
Aurass:2006, Kiener:2006, Hurford:2006, Mandrini:2007}. The
relationship of the Moreton wave to radio observations has already
been studied in \cite{Pick:2005}. In this paper, we focus on the
kinematical analysis of the Moreton wave, its relationship to the
flare, the CME, coronal dimmings and type II radio bursts, in order
to get insight into the wave characteristics and its initiating
agent.

\section{Data}

We analyzed the Moreton wave of 2003 October 28 and associated
phenomena (flare, EIT wave, coronal dimmings, type II radio burst)
using the following data sets.

\begin{enumerate}
  \item The Moreton wave is studied in  H$\alpha$ filtergrams recorded by the Meudon
Heliograph (France) which provides simultaneous observations of the
full Sun at three different wavelengths in the H${\alpha}$ spectral
lines (H${\alpha}$ line center, H${\alpha}$ $+$~0.5~$\hbox{\AA}$ and
H${\alpha}$ $-$~0.5~$\hbox{\AA}$) with an imaging cadence of $\sim$1
min.

\item The EIT wave and the coronal dimmings are studied in full-disk EUV
images by EIT/SoHO \citep{Delaboudiniere:1995}. The analysis is
conducted in the 195~$\hbox{\AA}$ bandpass which has a time cadence
of $12$~min. The coronal dimmings are also analyzed with TRACE
\citep[Transition Region And Coronal Explorer,][]{Handy:1999}.
High-resolution 195~$\hbox{\AA}$ filtergrams with a field-of-view
(FoV) of 380$\arcsec \times$340$\arcsec$ around the flare site are
available with a time cadence of $\approx$~8~sec.

\item The associated CME was observed by SoHO/LASCO
\citep[][]{Brueckner:1995}. We use the information from the LASCO
CME catalog at \verb"http://cdaw.gsfc.nasa.gov/cme_list/"
\citep{Yashiro:2004}.

\item The associated flare is studied in soft X-rays (SXRs) by the GOES10
satellite and in hard X-rays (HXRs) $>$150~keV provided by the
spectrometer SPI onboard INTEGRAL \citep[][]{Kiener:2006}.

\item The associated type II radio burst is analyzed utilizing the dynamic radio spectrum recorded by the radio
spectrographs of the Astrophysikalisches Institut Potsdam
\citep[AIP;][]{Mann:1992}. Positions of the radio source are derived
from observations of the Nan{\c c}ay Radioheliograph
\citep[NRH;][]{Kerdraon:1997}.

\item A full-disk magnetogram recorded by SoHO/MDI (at 11:00:03~UT; pixel size $\approx$~2$\arcsec$) is used to study the magnetic context of the event \citep{Scherrer:1995}.
\end{enumerate}

\section{Results}

\subsection{Event overview}
\begin{figure}
    \centering
    \resizebox{1\hsize}{!}{\includegraphics{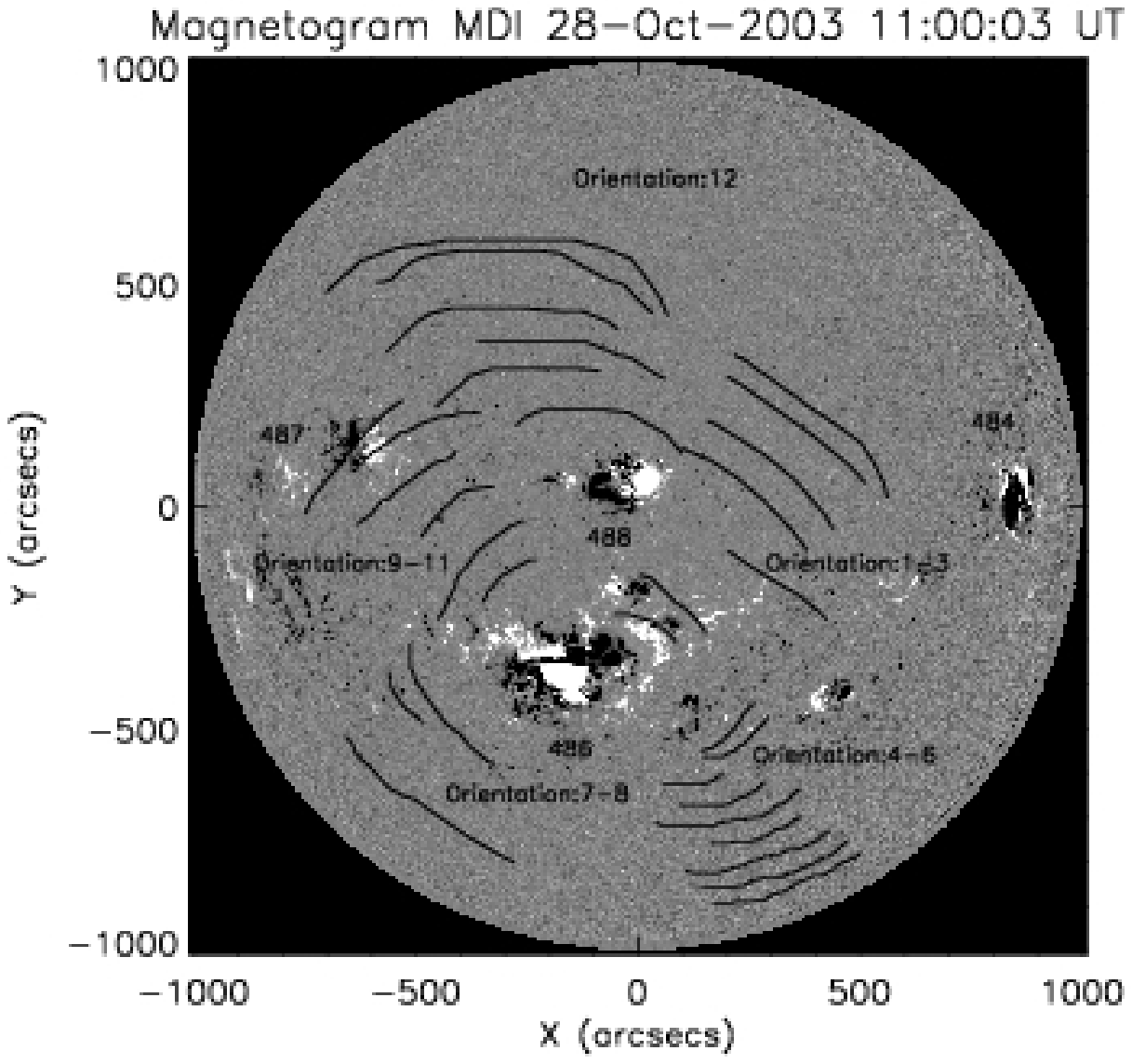}}
    \caption{Leading edges of all wavefronts visible during the time interval 11:02 -- 11:13~UT plotted on a
MDI/SoHO magnetogram recorded at 11:00~UT. Active region (AR)
numbers and the five sectors, in which the Moreton wavefronts were
propagating, are indicated.} \label{img:plot_magnetogramAR}
    \end{figure}

The Moreton wave under study was launched during a powerful
flare/CME event which occurred in NOAA AR10486 (S16$\degr$,
E08$\degr$) on October 28, 2003. NOAA AR10486 had a complex magnetic
configuration of $\beta \gamma \delta$ and was surrounded by several
other large and complex ARs (e.g. AR10484, AR10488; see
Fig.~\ref{img:plot_magnetogramAR}). The time range between
19-Oct-2003 and 4-Nov-2003 was characterized by an extremely high
level of solar activity during which 12 X-class flares occurred. On
28-Oct-2003 AR10486 produced a X17.2/4B two-ribbon flare. The GOES10
SXR flux showed the flare onset in the 1$-$8~$\hbox{\AA}$ channel at
$\approx$11:01~UT reaching peak at 11:10~UT. The INTEGRAL HXR
observations cover the total flare impulsive phase which lasted
roughly 15~min. The increase of the HXR flux ($>$150~keV) started at
11:02:00~UT and peaked for the first time at 11:02:40~UT. The
associated fast halo CME had a mean plane of sky velocity
$v\approx$~2500~km~s$^{-1}$ and its linear back-extrapolated launch
time is 11:01~UT (LASCO CME catalogue, \cite{Yashiro:2004}). This
goes along with a filament activation in the
southeastern quadrant at 11:01~UT. \\

\begin{figure*}
\centering \resizebox{0.85\hsize}{!}{\includegraphics{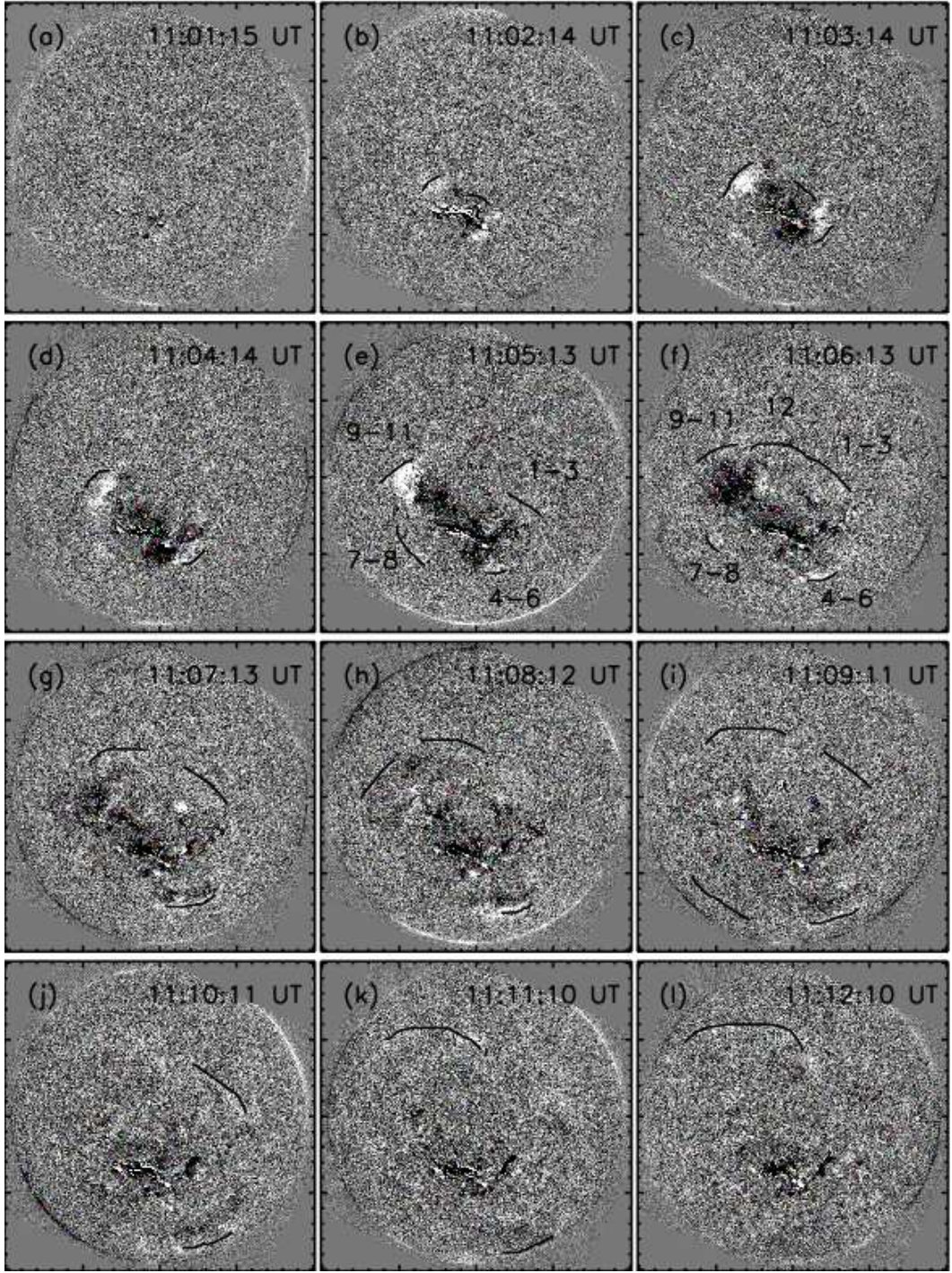}}
\caption{Sequence of H$\alpha$ blue-minus-red wing images after subtraction of a pre-event reference frame. 
The identified leading edges of the Moreton wavefronts (seen as
bright fronts) are indicated by black lines for all different
propagation directions. The corresponding sectors are indicated in
images (e) and (f). Each image shows a FoV of 2000$\arcsec$ $\times$
2000$\arcsec$ around Sun center.} \label{img:meudon_wavedifffullsun}
\end{figure*}

The Moreton wave can be identified during the time interval
11:02--11:13~UT. The kinematics of the Moreton wave is analyzed
applying two different methods. The first method relies on the
visual determination of wavefronts in a series of difference images.
The second one is based on the analysis of intensity profiles,
derived along the wave fronts, so-called perturbation profiles.

\subsection{Visual method}\label{subsec:vismethod}

We subtracted red and blue wing filtergrams recorded simultaneously
at H$\alpha  + 0.5 \hbox{\AA}$ and H$\alpha  - 0.5 \hbox{\AA}$ with
the Meudon Heliograph. These images contain information on the
plasma line-of-sight velocity (``dopplergrams''). In order to
maximize the wave contrast, we produced difference images by further
subtracting a blue-minus-red pre-event image (11:00~UT).
Figure~\ref{img:meudon_wavedifffullsun} shows this dopplergram
sequence. The Moreton wave appears as a bright, arc-like transient
disturbance, propagating away from the flare site in different
directions, spanning almost over the complete solar disk. We studied
the wave kinematics separately for four different directions, in
which the wave could be distinctly followed. The four sectors 1$-$3,
4$-$6, 7$-$8, 9$-$11 (denoted by hours on the clock-face) and the
visually identified wavefronts are indicated in
Fig.~\ref{img:meudon_wavedifffullsun}. We note that the wavefronts
in direction 9$-$11 were affected by refraction and reflection at
the strong magnetic fields around AR 10488, which resulted in the
wave propagation towards North after 11:06~UT (denoted as sector 12
in Fig.~\ref{img:meudon_wavedifffullsun}). The wave behavior in
sector 12 is not considered in the following analysis.

\begin{figure}
    \centering
     \resizebox{1\hsize}{!}{\includegraphics{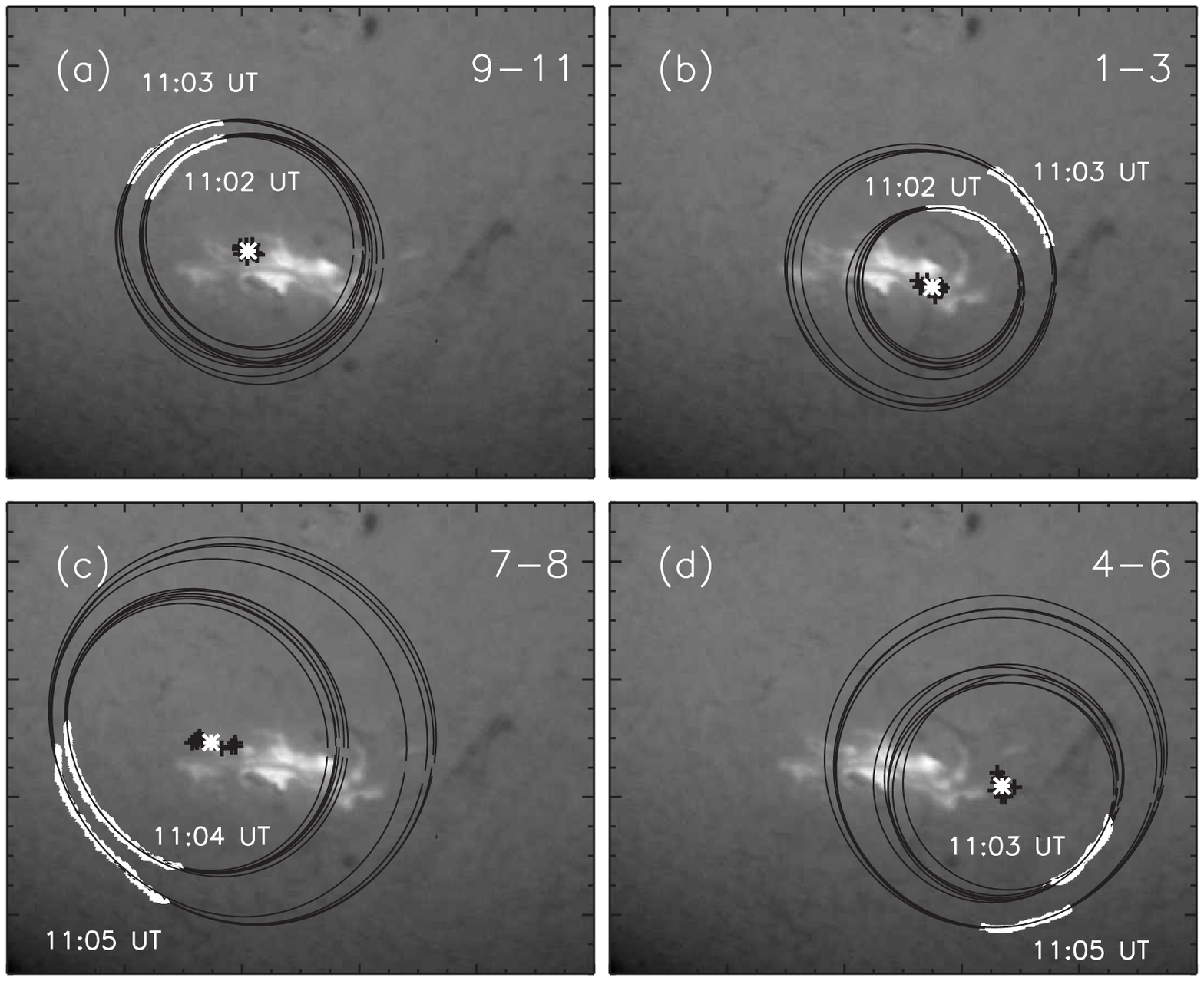}}
    \caption{H${\alpha}$$-$0.5~$\hbox{\AA}$ image at 11:12~UT. The visually determined leading edges
    of the first and second visible
    wavefronts (marked as white lines) for the sectors (a) 9$-$11, (b) 1$-$3, (c) 7$-$8, (d) 4$-$6 are overlayed. The
    wavefronts were drawn several times in order to estimate errors (first wavefront: six times, second wavefront: four times).
    The superposed
    ellipses (black) correspond to fitted circles in the deprojected heliographic 2D plane, and are
    used for calculating the wave radiant points (black pluses) and their mean values (white crosses). The
    plotted FoV is $x$=[$-$600$\arcsec$, +400$\arcsec$], $y$=[$-$700$\arcsec$, +100$\arcsec$] with
    the origin at the solar disk center.} \label{img:waveellorzoominblackred}
\end{figure}

The Moreton wave ``radiant point'' (source center) was estimated by
applying circular fits to the earliest observed wavefronts for each
propagation direction (see Fig.~\ref{img:waveellorzoominblackred}),
whereby the projection effect due to the spherical solar surface was
taken into account \citep[following][]{Warmuth:2004,Veronig:2006}.
In order to estimate the error of the determined wave center, each
of the earliest wavefronts was drawn for six times (first
appearance) or four times (second appearance). The determined
wave centers for the four directions are:\\
$[x,y]\approx[-253\arcsec\pm31\arcsec,-307\arcsec\pm6\arcsec]$ for direction 7--8, \\
$[x,y]\approx[-190\arcsec\pm11\arcsec,-315\arcsec\pm7\arcsec]$ for direction 9--11, \\
$[x,y]\approx[-51\arcsec\pm16\arcsec,-376\arcsec\pm8\arcsec]$ for direction 1--3, and \\
$[x,y]\approx[+69\arcsec\pm~9\arcsec,-381\arcsec\pm14\arcsec]$ for direction 4--6. \\
The wave centers derived for the four propagation directions can be
grouped into two significantly different radiant points at the mean
coordinates
$[x,y]\approx[-221\arcsec\pm39\arcsec,-311\arcsec\pm8\arcsec]$ and
$[x,y]\approx[9\arcsec\pm61\arcsec,-379\arcsec\pm11\arcsec]$, on the
East and West borders of the source AR (see
Fig.~\ref{img:waveellorzoominblackred}). This suggests that the wave
propagation to the East (7--8 and 9--11) and to the West (1--3 and
4--6) can be attributed to two different source sites. Accordingly,
the wave kinematics are derived using two different source site
coordinates.

\begin{figure}
\centering \resizebox{1\hsize}{!}{\includegraphics{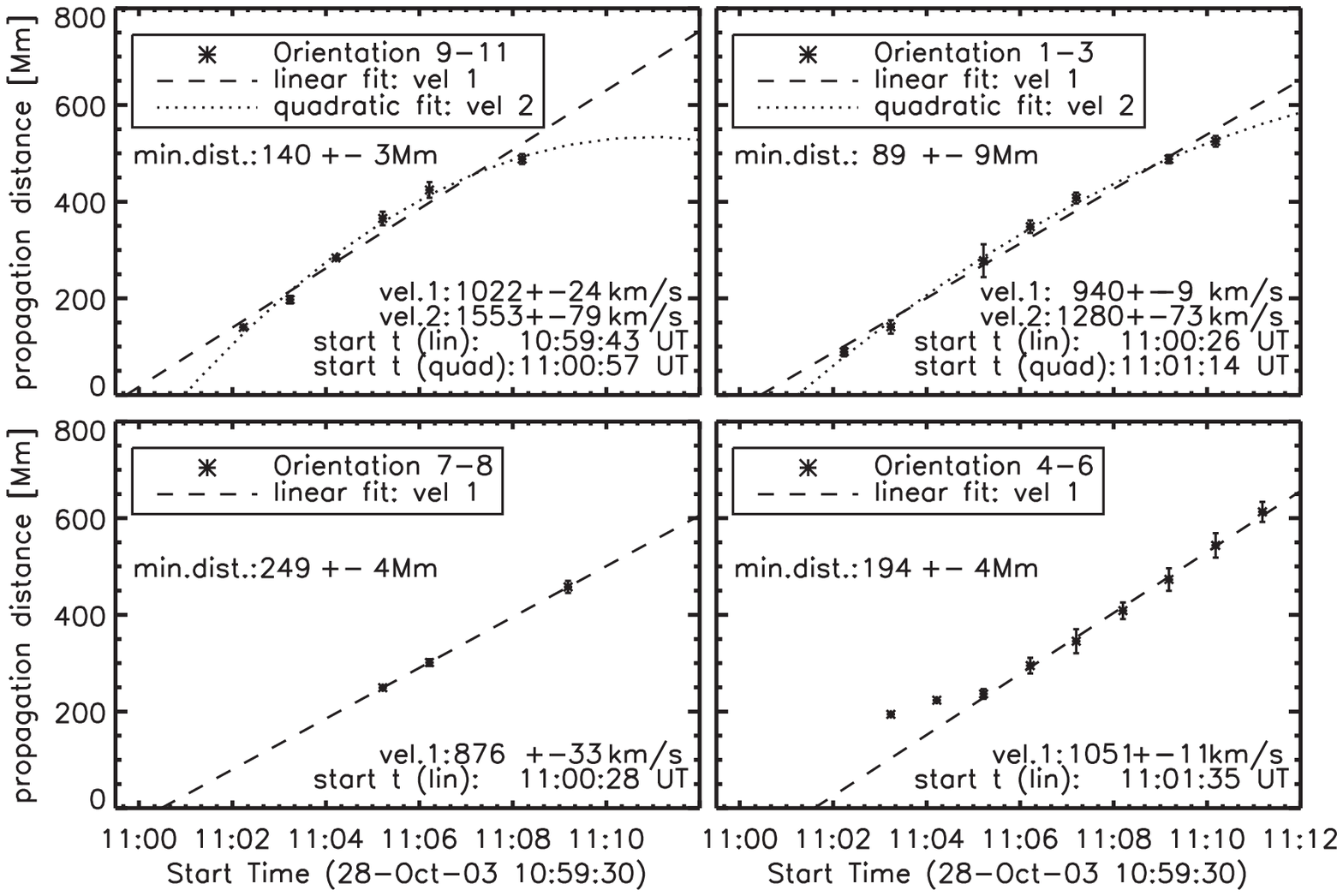}}
\caption{Kinematics of the Moreton wave. Distance vs.\ time diagrams
derived separately for four different propagation directions (cf.\
Fig.~\ref{img:plot_magnetogramAR}). The dashed and dotted lines
indicate the linear and quadratic fits to the kinematical curves,
respectively. The first two data points for direction 4$-$6 were
excluded in the fitting process, since they are uncertain due to the
wave passage over a filament. Note that the error bars do not
reflect uncertainties on the derived distances but deviations of the
visually drawn wavefronts from the extrapolated circular shape in
the deprojected heliographic 2D plane.}
\label{img:velocityallresult}
\end{figure}

To derive the wave kinematics, we calculated for each point of the
wavefronts its distance from the respective radiant point along
great circles on the solar surface, and then averaged over all
points of the wavefront. Figure~\ref{img:velocityallresult} shows
the distance-time diagrams for the four different propagation
directions. The wave was first visible at a distance of
$\approx$90~Mm from the wave's radiant point in direction 1$-$3, and
can be followed up to distances of 500--600~Mm in the different
directions. In direction 4$-$6, the winking of a filament can be
observed after the wave passage ($\approx$11:04~UT), which makes the
wavefront determination difficult for this phase.

Linear and quadratic functions are fitted to the derived kinematical
curves of the wave (see Fig.~\ref{img:velocityallresult}), from
which we extrapolated the start times, ranging from 10:59:40 to
11:01:40~UT. The mean velocities derived from the linear fits are
940$\pm$10 km~s$^{-1}$, 1050$\pm$10 km~s$^{-1}$, 880$\pm$30
km~s$^{-1}$, and 1020$\pm$20 km~s$^{-1}$ for the orientations 1$-$3,
4$-$6, 7$-$8, and 9$-$11, respectively. The quadratic fits indicate
wave deceleration for directions 1$-$3 and 9$-$11 with
--1420$\pm$360~m~s$^{-2}$ and --2960$\pm$280~m~s$^{-2}$,
respectively.

\subsection{Perturbation profiles}

In the second approach, the Moreton wave propagation was followed by
analyzing the perturbation intensity-profiles $\Delta I(r,t)$. It
has to be noted that the perturbation profile analysis can only be
carried out if the disturbance is strong enough \citep[see examples
in][]{Warmuth:2004b}. Moreton waves are in most cases too faint and
irregular to be analyzed by this method. We also note that there are
simplifying assumptions involved in the profile method: It is
supposed that the ambient
medium is homogeneous and the wavefronts expand circularly.\\

First of all, a specific sector is chosen for which the intensity
evolution is calculated. Then the intensity values are averaged over
annuli with a chosen angular width and a thickness of 5~Mm. This
results in a mean intensity as a function of distance $r$ (measured
along the solar chromosphere) from the wave's radiant point. The
procedure is repeated for each filtergram, until the Moreton wave
fades away. Since we investigate intensity changes $\Delta I$, we
use the blue-minus-red wing base-difference images, where the wave
contrast is best.

\begin{figure}
\centering \resizebox{0.8\hsize}{!}{\includegraphics{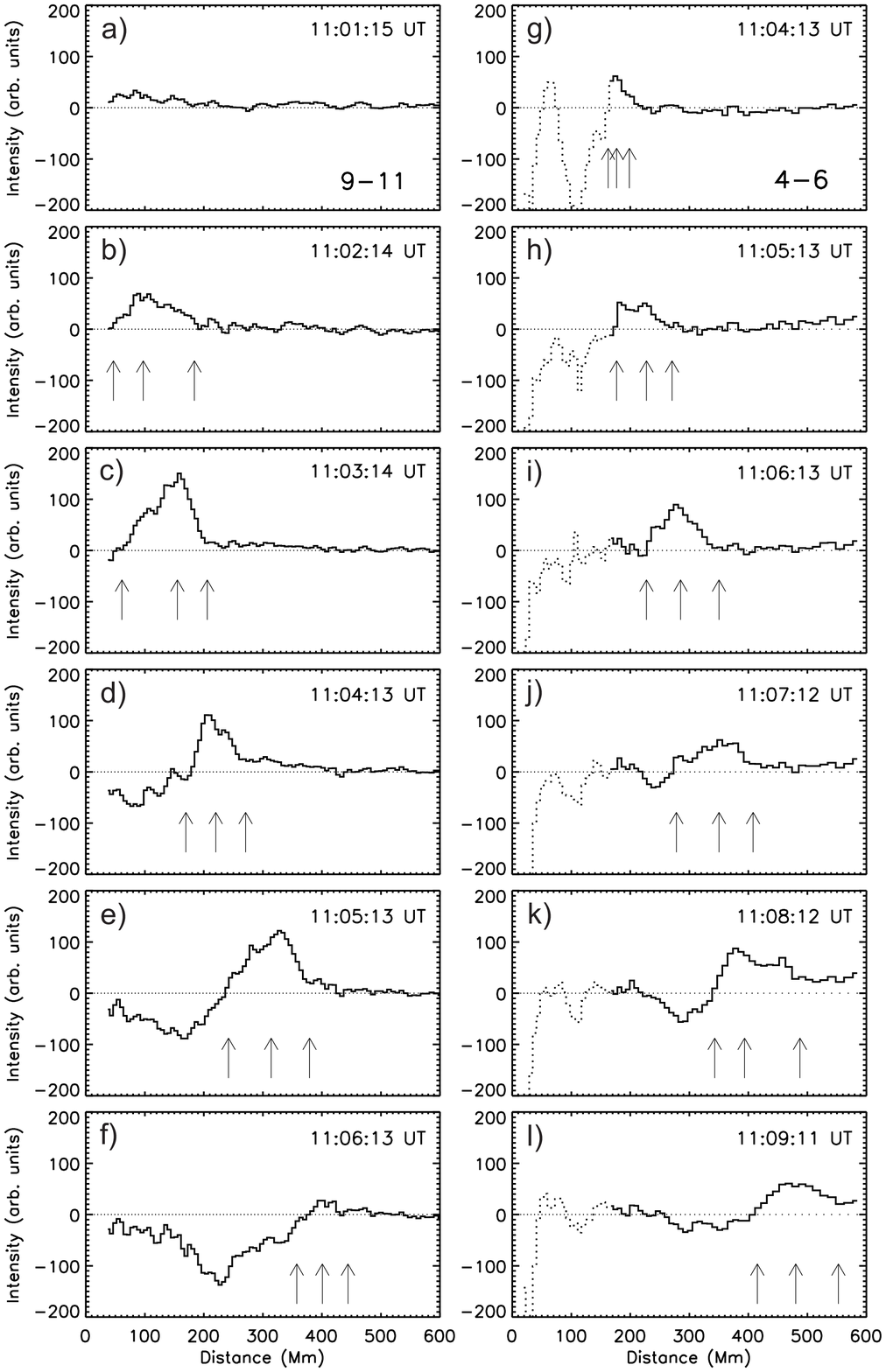}}
\caption{Perturbation profiles derived for the wave segments
propagating in sectors 9--11 and 4--6. Arrows indicate the positions
of the leading edge, the peak amplitude, and the trailing edge,
respectively.} \label{img:rb9-11rb4-6}
\end{figure}

The method was carried out for the directions 9$-$11 and 4$-$6
because of their high intensities and their rather constant sector
spans. However, the wavefronts become more and more irregular and
show noticeable changes in their shape and propagation direction.
Thus, with the profile method we can track the wave over shorter
distances than with the visual method. The advantage of this method
is that it provides us with more information on the perturbation
characteristics. Figure~\ref{img:rb9-11rb4-6} displays the
perturbation profiles of the propagating wavefronts. From these
profiles, the locations of the leading edge $r_l$, the trailing edge
$r_t$ and the peak amplitude $r_m$ are extracted (see arrows in
Fig.~\ref{img:rb9-11rb4-6}). The difference $\Delta r= r_l-r_t$
gives the width of the propagating disturbance.

\begin{figure}
\centering \resizebox{0.9\hsize}{!}{\includegraphics{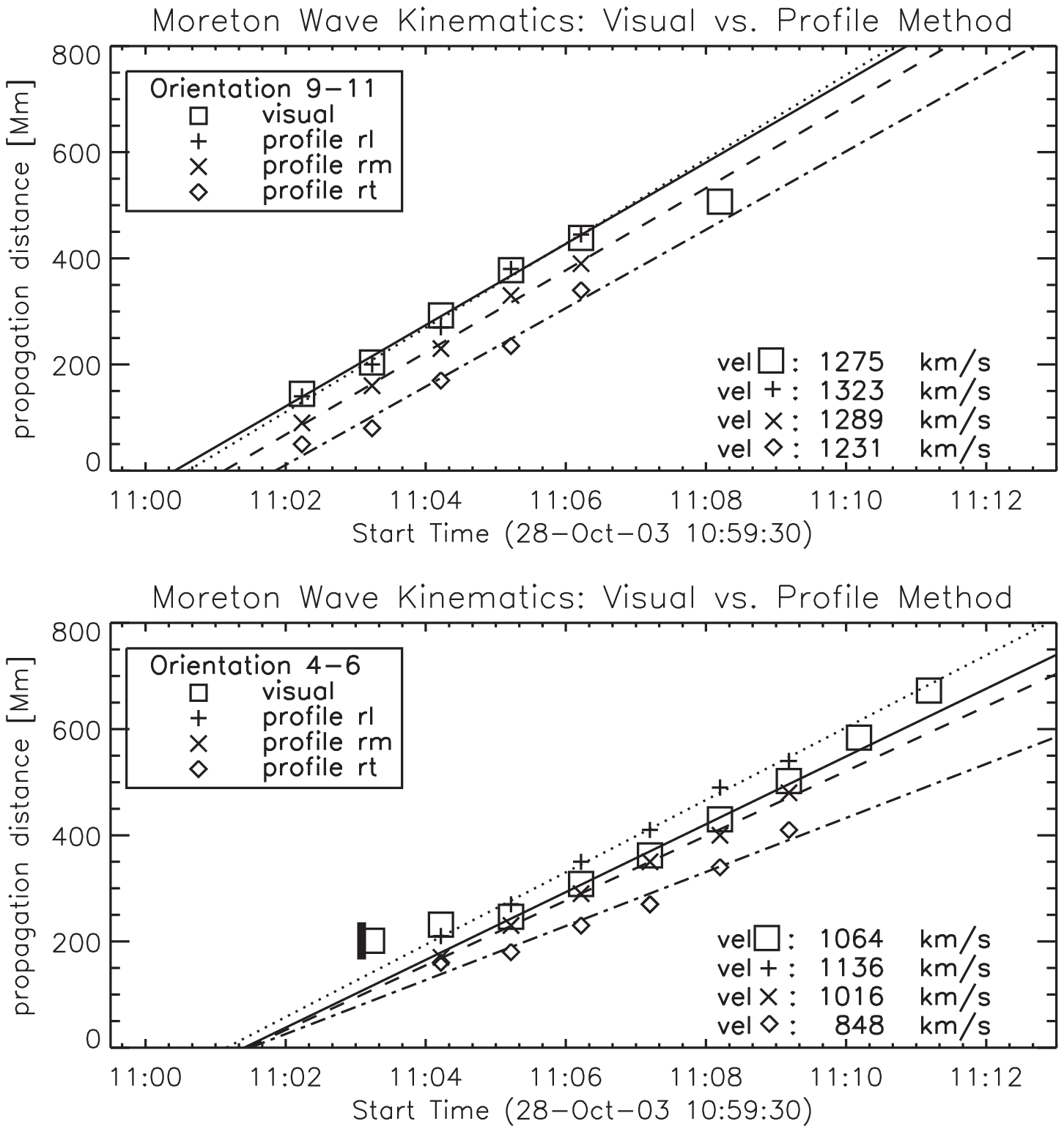}}
\caption{Kinematics of the Moreton wave (distance-time diagrams)
derived from the perturbation profiles for orientation 9$-$11 and
4$-$6 (cf. Fig.~\ref{img:rb9-11rb4-6}). The pluses indicate the
position of the leading edge $r_l$, the crosses the position of the
peak amplitude $r_m$, and the diamonds the position of the trailing
edge $r_t$. For comparison, we also plot the kinematics derived from
the visual method (squares). The solid, dotted, dashed and
dash-dotted lines indicate the linear fits to the data sets,
respectively. The annotations vel~$\Box$, vel~$+$, vel~$\times$ and
vel~$\diamond$ refer to the velocity values derived from the linear
fits to the different time-distance data. The thick black vertical
bar in direction 4--6 indicates the position of the type II burst
within the measurement error (cf. Sect. 3.6).}
\label{img:intvelocitylast}
\end{figure}

From the calculated perturbation profiles, we derived the wave
kinematics. Figure~\ref{img:intvelocitylast} shows the time-distance
plots of the Moreton wave derived from the profile method, in
comparison with the curves derived from the visual method. The
extrapolated start times lie in the range of 11:00:40--11:01:50~UT,
and correspond well to the starting times derived from the visual
method. For the direction 9$-$11, the mean velocities derived from
the linear fits are 1320~km~s$^{-1}$, 1270~km~s$^{-1}$ and
1230~km~s$^{-1}$ for $r_l$, $r_m$ and $r_t$, respectively. The
values for orientation 4$-$6 are 1140~km~s$^{-1}$, 1040~km~s$^{-1}$
and 850~km~s$^{-1}$. The velocities derived by the profile method
differ from those derived from the visual-method. However, this
difference is not a result from the different methods, but merely
results from the fact that with the profile method the late-phase
evolution of the wave, when it has already decelerated, cannot be
extracted. Using an identical time range for the fitting procedure
for both methods (orientation 9--11: 11:02~UT -- 11:06~UT;
orientation 4--6: 11:04~UT -- 11:09~UT), we obtain from the visual
method mean velocities of 1280~km~s$^{-1}$ and 1070~km~s$^{-1}$,
which is consistent with the velocities obtained from the profile
method (peak amplitude) for the two sectors (cf.\
Fig.~\ref{img:intvelocitylast}).

Figure~\ref{img:intvelocitylast} also indicates that the distance
between the leading edge $r_l$, the peak amplitude $r_m$ and the
trailing edge $r_t$ increases with time, which is especially
prominent in direction 4--6. The peak amplitude of the perturbation
increases in the beginning (see Fig.~\ref{img:rb9-11rb4-6}, panels~g
and h), and after achieving maximum (panel~i), subsequently
decreases (panels~j to l). In direction 9--11, the wavefront
steepens from panel a to c. We emphasize that panel c (11:03:14~UT)
shows the situation close to the first appearance of the type II
burst source (see Sect.~\ref{seq:type2}), which reveals the
formation of the coronal shock. The dip in the profile (at the
trailing part of the wave) is a signature of the upward relaxation
of the compressed chromospheric plasma. It is not visible in the
beginning but becomes more and more remarkable as the wave evolves.

\begin{figure}
\centering \resizebox{0.9\hsize}{!}{\includegraphics{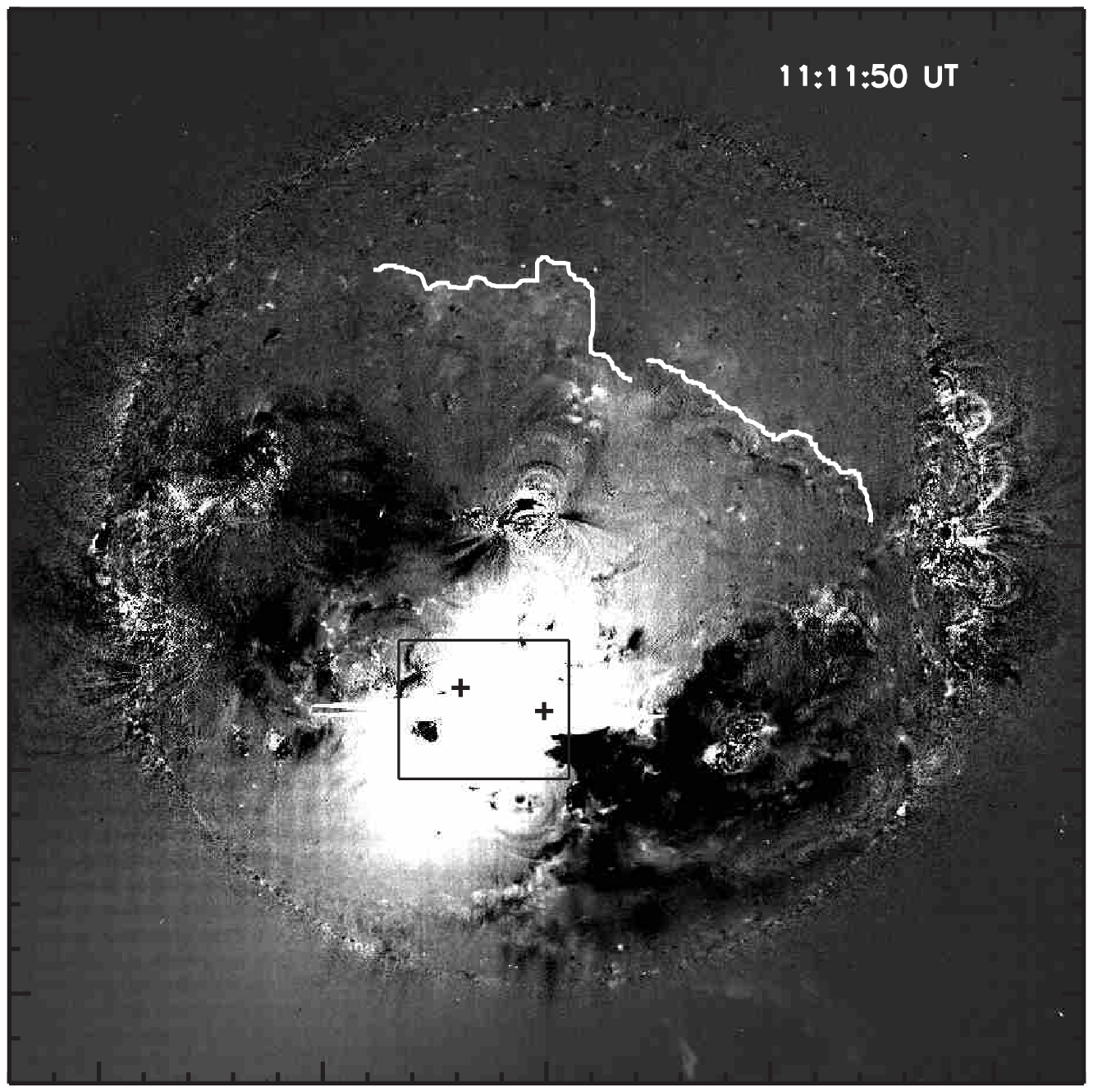}}
\caption{EIT base-difference image of 11:11:50~UT (reference image
is 9:24:50~UT). The leading edge of the EIT wavefront is indicated
by a white line. Coronal dimmings can be observed on both sides of
the flare site (locally and globally). The black pluses show the two
radiant points of the Moreton wave. The rectangle indicates the
TRACE field of view (cf.~Figure~\ref{img:dimming_overview_8panel}).
The image shows a FoV of 2000$\arcsec$ $\times$ 2000$\arcsec$.}
\label{img:eit_basediff924blackwave}
\end{figure}

\subsection{EIT wave}

\begin{figure}
\centering \resizebox{1\hsize}{!}{\includegraphics{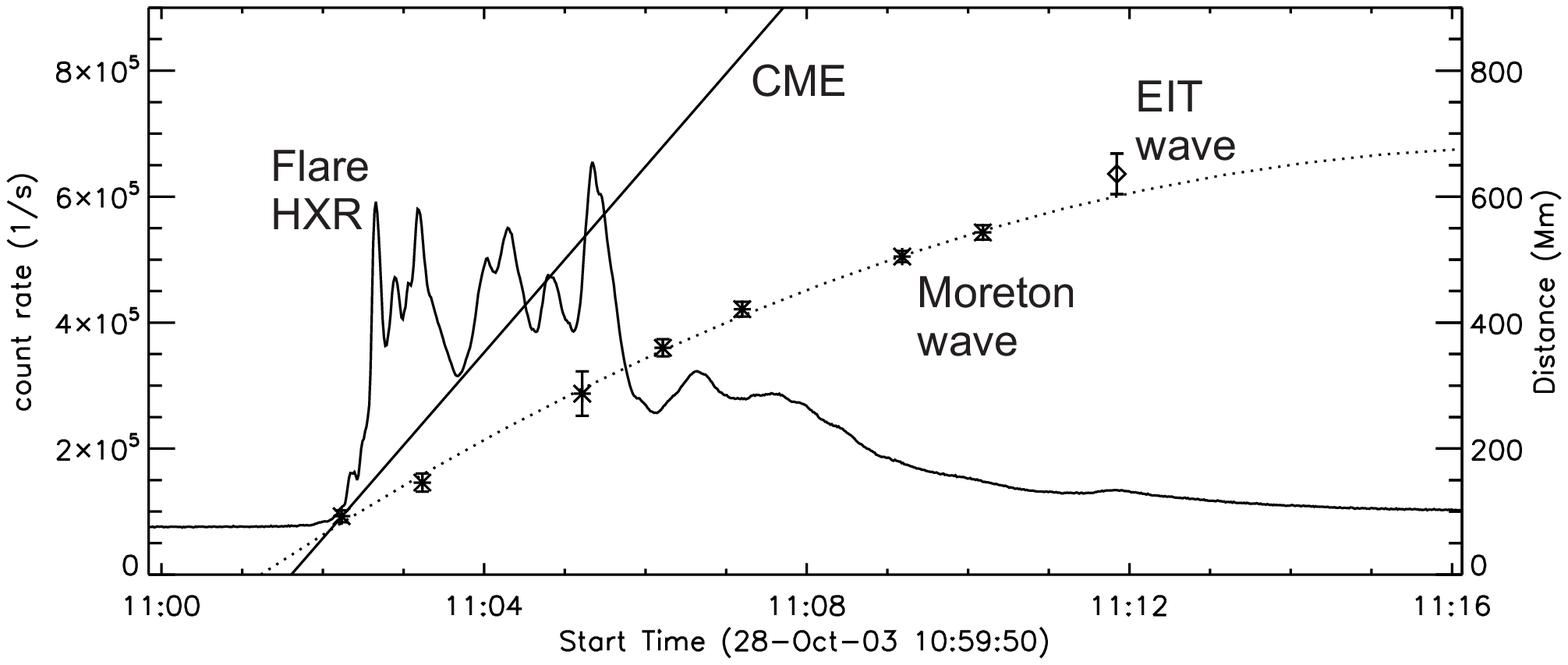}}
\caption{Kinematics of the Moreton wavefronts in direction 1$-$3
(visual method; displayed by asterisks) together with the second
order fit, the position of the EIT wavefront at 11:12~UT in
direction 1--3 (displayed by diamond), the linear back-extrapolated
CME height-time curve, and the flare HXR flux above 150 keV.}
\label{img:plot_HXR}
\end{figure}

EIT images provide information on large-scale coronal changes associated  
with flare/CME events. Figure~\ref{img:eit_basediff924blackwave}
shows a base-difference EIT 195~$\hbox{\AA}$ image at 11:11:50~UT
(reference image is 9:24:50~UT). Due to the 12~min time cadence of
the EIT instrument, the transient EIT wave (bright diffuse arcs
toward North and Northwest of AR~10488) is observable in only one
single frame at 11:11:50~UT in two sectors. It has already traveled
across a significant portion of the solar disk. Since the EIT wave
can only be traced in one single image, no kinematics can be
derived. However, the position of the EIT wavefront can be compared
to the Moreton wavefronts observed in H$\alpha$
(Fig.~\ref{img:plot_HXR}).

Figure~\ref{img:plot_HXR} shows a combined plot of the Moreton wave
kinematics (direction 1--3; visual method), the flare HXR flux
measured by INTEGRAL, the back-extrapolated CME propagation curve
(from the LASCO catalog) and the location of the EIT 
wavefront (direction 1--3). Extrapolating the 2nd order fit to the
Moreton wave distance-time diagram in this direction, we find that
the EIT wave is slightly ahead of the Moreton wave but basically on
the same kinematical curve. The fact, that the EIT wavefront is
located some 10~Mm ahead of the corresponding Moreton wave was also
reported from other studies \citep[e.g.][]{Warmuth:2004,
Veronig:2006}. Assuming that both are related phenomena, it can be
explained by the different heights of the Moreton wave
(chromosphere) and EIT wave (lower corona) observations and the
inclination of the expanding coronal wave dome as envisaged in
Uchida's theory.

\subsection{Coronal Dimmings}

\begin{figure}
\centering
  \resizebox{0.7\hsize}{!}{\includegraphics{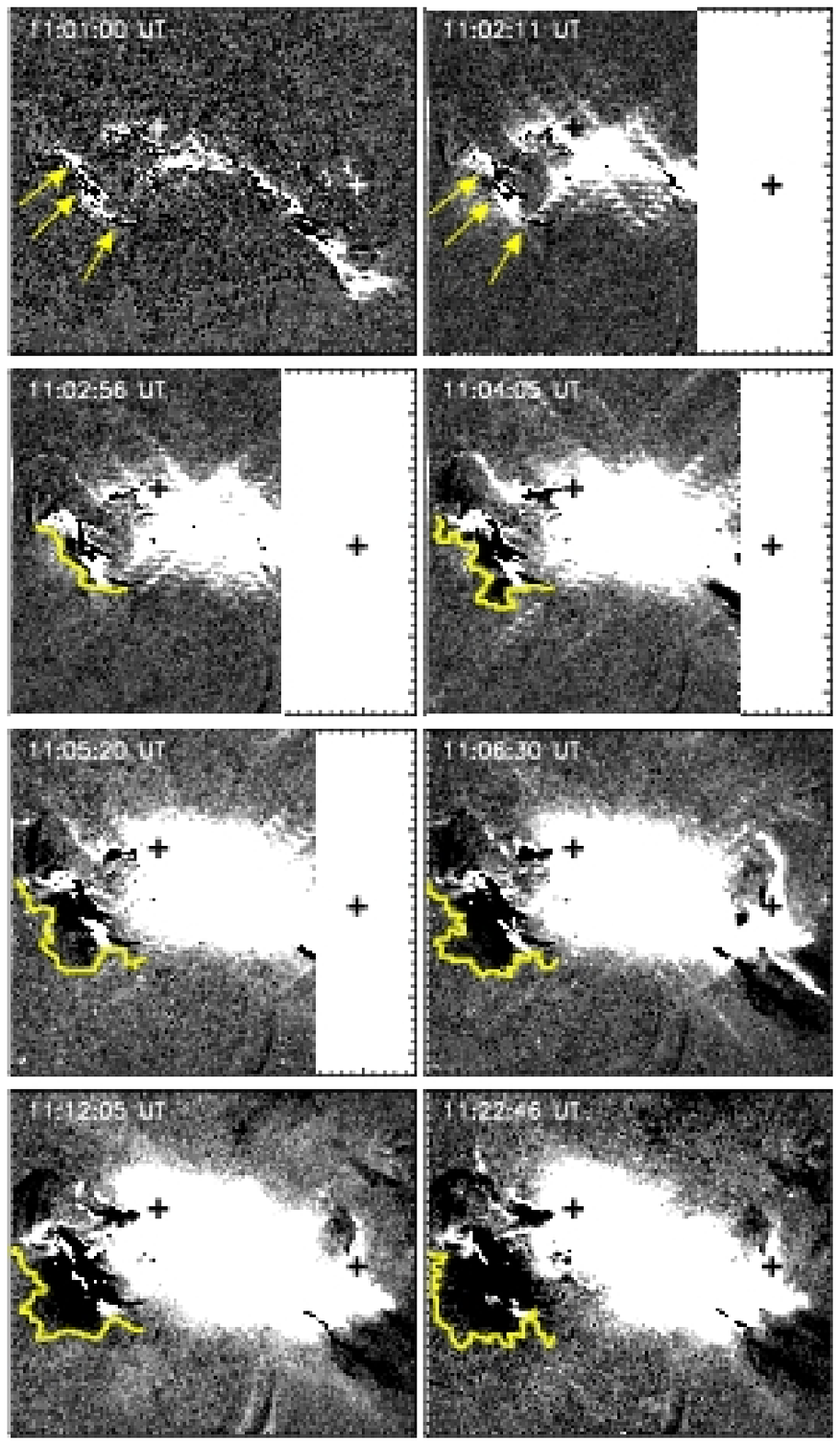}}
  \caption{Evolution of the double coronal dimming region (black areas at East and West edges of the flare site).
  The pluses show the two derived radiant points of the Moreton wave. The
    plotted FoV is $380 \arcsec \times 340\arcsec$ centered around [x,y]=$[-140\arcsec, -380\arcsec]$.
    The arrows and lines indicate the outer edge of the dimming region.}
  \label{img:dimming_overview_8panel}
\end{figure}

Bipolar coronal dimmings are believed to be the footprints of
large-scale flux-rope ejections \citep[e.g.][]{Hudson:1997,
Webb:2000}. In the event under study, coronal dimmings are
detectable in EIT~195~$\hbox{\AA}$ (see
Fig.~\ref{img:eit_basediff924blackwave}) as well as
TRACE~195~$\hbox{\AA}$ (see Fig.~\ref{img:dimming_overview_8panel}).
EIT observations are available as full-disk images with a time
cadence of $\approx$~12~min. TRACE observations are available for a
FoV of 380$\arcsec\times$340$\arcsec$ around the flare site with a
high time cadence ($\approx$~8~sec) during 10:14~UT and 11:45~UT.
Figure~\ref{img:eit_basediff924blackwave} shows an EIT~195~{\AA}
base-difference image. It displays the global coronal dimmings,
which are expanding over the whole hemisphere indicative of a
complete restructuring of the corona \citep[e.g.][]{Zhukov:2007}. In
addition to these global or ``secondary'' dimmings
\citep[][]{Mandrini:2007}, localized bipolar dimmings in the near
vicinity of the flare site are also observed. These ``core
dimmings'', as quoted by \citet{Mandrini:2007} are difficult to
observe in EIT but are well captured with TRACE. Here, we focus on
these small, bipolar dimming regions and their evolution.

Figure~\ref{img:dimming_overview_8panel} shows a sequence of
TRACE~195~$\hbox{\AA}$ base-difference images. A pre-event image
recorded at 10:59:26~UT was subtracted from the whole image series.
At 11:00:29~UT we can observe the first small brightenings along the
neutral line in AR10486.
The first signature of the bipolar dimmings, which are observed as
localized dark regions on each side of the flare site, can be
detected at 11:01:00~UT (see
Fig.~\ref{img:dimming_overview_8panel}). They lie on opposite
east-west edges of the source AR and probably reflect the two CME
expanding flanks. We also note that they are roughly located on the
same axis as the two radiant centers of the wave. The time of first
appearance of these bipolar dimming corresponds to the CME onset,
which is determined from the linear back-extrapolation of the CME
time-distance curve to be at $\approx$11:01~UT.

\begin{figure}
  \resizebox{1\hsize}{!}{\includegraphics{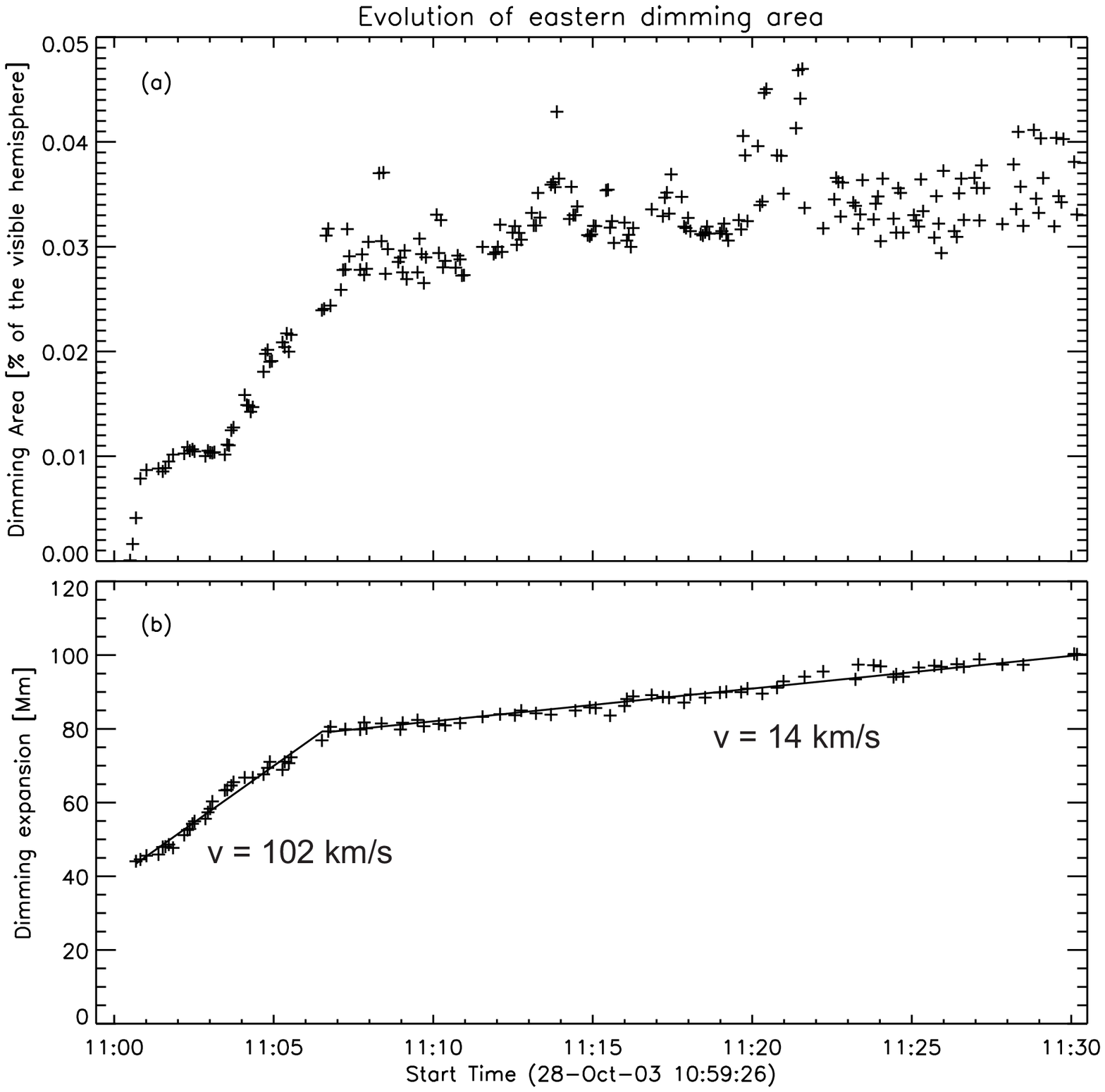}}
  \caption{(a) Evolution of the area (in \% of the visible solar hemisphere) occupied by the eastern coronal dimming region as measured from TRACE~195~$\hbox{\AA}$ images.
  (b) Kinematics of the outer border of the eastern coronal dimming region. Linear fits are applied separately to the early ($<$11:06~UT) and
  later evolution phase of the dimming.}
  \label{img:dimming_evolution_left}
\end{figure}

In the high cadence TRACE EUV images, we can study the evolution of
the bipolar dimmings. Since the Western dimming region was only
partially covered by the TRACE FoV (cf.\
Fig.~\ref{img:dimming_overview_8panel}), we concentrate here only on
the Eastern dimming region. To obtain information on the area of the
bipolar dimming regions, we used specific intensity levels and
summed up the enclosed area (given in percentage of the entire solar
disk; see Fig~\ref{img:dimming_evolution_left}a). 

The expansion velocity of the dimming regions (presumably related to
the CME flanks) is derived by measuring the propagation of the outer
edge of the dimming region, indicated by grey lines in
Fig.~\ref{img:dimming_overview_8panel}.
Figure~\ref{img:dimming_evolution_left}b shows the distance-time
diagram for the kinematics of the eastern dimming region. Due to the
distinctive change in its propagation velocity, we applied two
separate linear fits to the kinematical curve. The dimming shows a
fast expansion ($v \approx 102$ km~s$^{-1}$) until 11:06~UT,
thereafter it drastically slows down by almost an order of magnitude
to $v \approx 14$~km~s$^{-1}$.  We note that
Fig.~\ref{img:dimming_evolution_left}a and
Fig.~\ref{img:dimming_evolution_left}b show a similar evolution with
fast growth up to $\approx$11:06~UT and much slower changes
thereafter. At the time of the fast growth, the CME is in its
initial phase.

\subsection{Type II radio burst}\label{seq:type2}
According to \cite{Uchida:1968}, Moreton waves appear at the
intersection line of an expanding, coronal fast-mode shock and the
chromosphere. Another shock signature are type II radio bursts
\citep{Uchida:1974}, excited by electrons accelerated at the shock
front. Type II bursts usually show the fundamental and harmonic
emission band, both frequently being split in two parallel lanes,
so-called band-split \citep{Nelson:1985}. The interpretation of the
band split as the plasma emission from the upstream and downstream
shock regions, was affirmed by \citet{Vrsnak:2001}. Thus the
band-split can be used to obtain an estimate of the density jump at
the shock front \citep{Vrsnak:2002, Magdalenic:2002}. From that, one
can infer the pressure jump, which governs the downward compression
of the chromosphere after the shock-front passage.

\begin{figure}
   \resizebox{1\hsize}{!}{\includegraphics{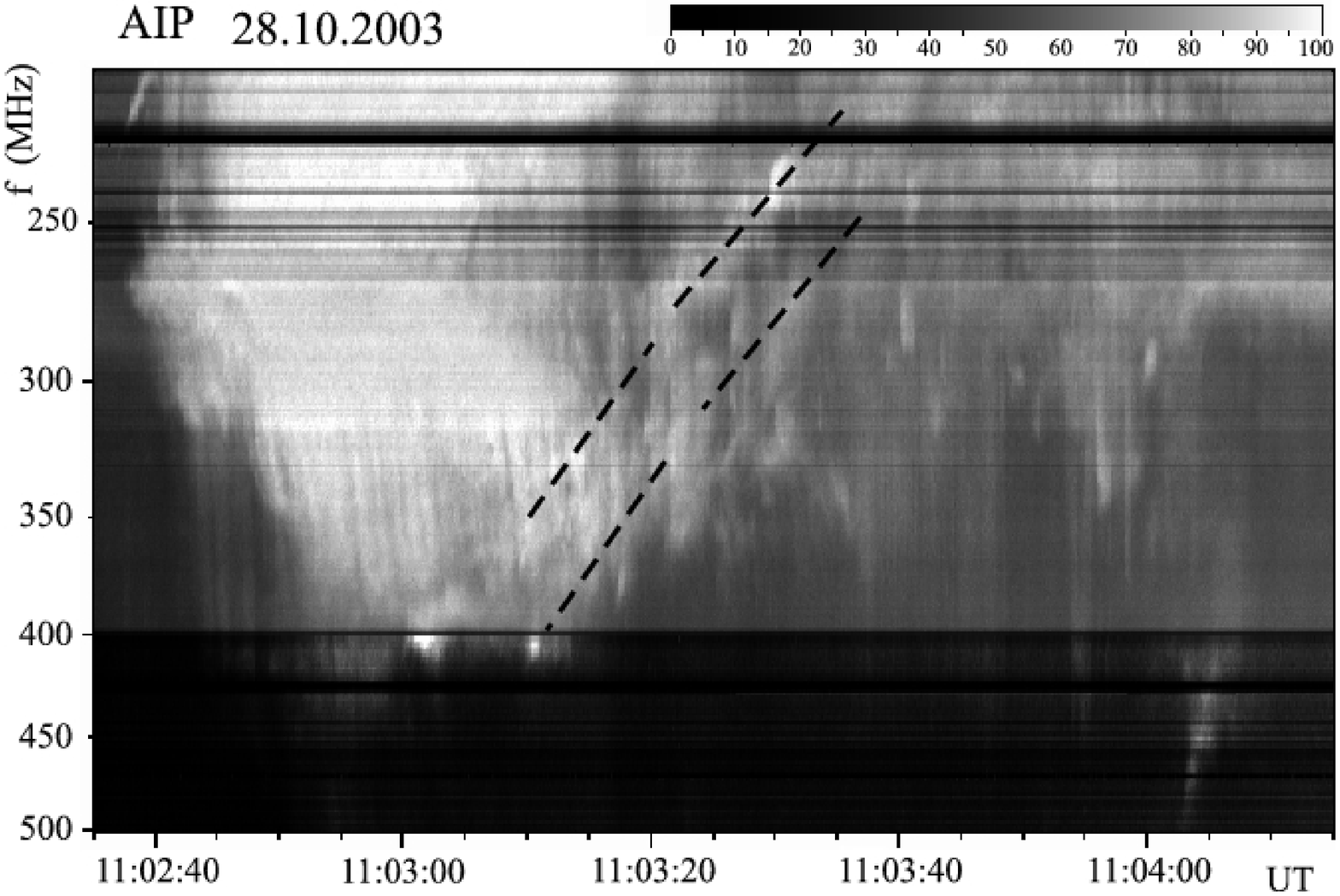}}
  \caption{Dynamic radio spectrum recorded by AIP spectrographs, displaying
  the high-frequency part of the metric type II burst. The two dashed lines indicate
  the band split of the type II harmonic lane. The metric type II continues
  to decameter wavelengths.}
  \label{img:osra_spect}
\end{figure}

We note that several aspects of the radio observations of this event
were already analyzed by \citet{Klassen:2005}, \citet{Pick:2005} and
\citet{Aurass:2006}.  The AIP dynamic radio spectrum
(Fig.~\ref{img:osra_spect}) shows complex and intense radio
emission, consisting of a group of type III bursts, a type II bursts
\citep{Klassen:2005}, and a complex type IV burst. For a detailed
dynamic spectrum covering frequency range 20 - 800 MHz, we refer to
Fig.~4 in \citet{Pick:2005}.

In the present study we focus on the type II burst that started
around 11:03 UT, at a frequency of 420 MHz. The back-extrapolation
of the emission lanes point to 11:02:30~UT, i.e., close to the first
peak of the HXR burst, and the intense type III bursts which
indicate the impulsive phase of the flare. This is also close to the
steepening phase of the Moreton wavefront of direction 9--11 (see
Fig.~\ref{img:rb9-11rb4-6}, panel~3).

Nan{\c c}ay Radioheliograph measurements at 411~MHz at 11:03:01~UT
and 11:03:11~UT show that the type II burst source was located at
$\approx$ $(x,y)$=$(0.2~R_{\sun}, -0.5~R_{\sun})$. Comparing the
type II source location with the Moreton wave front, we find that
the position of the radio source is cospatial with the first Moreton
wave front in the direction 4$-$6. In Fig.~\ref{img:intvelocitylast}
the position of the radio source is marked by a thick vertical bar,
with the length of the bar corresponding to the measurement error.

The back-extrapolation of the type II burst to 11:02:30~UT indicates
that the disturbance, which developed into a shock, was probably not
launched before that time. On the other hand, we found the
back-extrapolated start time of the Moreton wave to the western
radiant point at 11:01:35~UT (see Fig.~\ref{img:velocityallresult}).
This indicates that the Moreton wave was not launched from a
point-like source but rather from an extended source. From
Fig.~\ref{img:intvelocitylast} we find that at 11:02:30~UT the
back-extrapolation of the Moreton wave trajectory was at a distance
of $d\approx$~60--80~Mm, which provides us with a lower limit for
the source region size. The upper limit for the source region is
given by the first appearance of the Moreton wave, which was at
$\approx$~200~Mm.

Due to the complexity and overlapping of various radio emissions it
is difficult to analyze in detail the high-frequency part of the
type II burst. Nevertheless, in the period 11:03:10 to 11:03:40~UT,
we were able to recognize a band-split pattern, with a relative
bandwidth of $BDW=\Delta f/f\approx 0.15$\,--\,0.25 (see
Fig.~\ref{img:osra_spect}). The relative bandwidth $BDW$ is
determined by the density jump at the shock front
$X=\rho_{2c}/\rho_{1c}$, where $\rho_{1c}$ and $\rho_{2c}$ are the
densities upstream and downstream of the shock front \citep[for
details see][]{Vrsnak:2001}. Since $BDW\equiv
(f_2-f_1)/f_1=\sqrt{\rho_{2c}/\rho_{1c}}-1$, we find
$X_c=1.32$\,--\,1.56, where the subscript ``c" stands for ``corona".

\section{Interpretation and discussion}

\subsection{Source site and driver}

We found two separate initiation centers for the Moreton wavefronts
on opposite east-west edges of the source AR~10486. This implies
that actually two waves were launched simultaneously. This argues
against an initiation by the flare pressure pulse: first, a
simultaneous launch of two flare pressure pulses is very unlikely;
second, at the time of the first flare hard X-ray peak
(11:02:40~UT), indicating the first episode of powerful flare energy
release, we already observed the Moreton wave, this means that if
initiated by the flare there would have been no time for the wave to
reach a large amplitude to be observable. Nevertheless, we note that
an initiation scenario caused by the flare cannot be completely
excluded (for instance, assuming that the wave was launched already
{\it before} the time of the first peak in the flare energy
release).

The most likely interpretation of the observations is that the
Moreton wave is caused by the expanding flanks of the CME, evidenced
by the observed bipolar dimmings on east-west edges of the flare
site. This is further strengthened by the observations of two
Moreton wave radiant points, which lie roughly on the same axis as
the bipolar dimmings. We also note that the initial acceleration of
the wave coincides with the fast expanding phase of the dimming
evolution with $v\approx$ 100~km~s$^{-1}$, which lasts until
11:06~UT (cf.\ Fig.~\ref{img:dimming_evolution_left}). This
corresponds also well with the end of the steepening phase
(presumably indicating the shock formation) of the Moreton wavefront
profiles (cf.\ Fig.~\ref{img:rb9-11rb4-6}).

We also emphasize that the Moreton wave is not launched from two
localized points but from two extended sources. This  result is
derived from the type II burst appearance revealing the shock
formation, which is first detectable at $\approx$11:03~UT at a
certain distance ($\approx$60~Mm) from the Moreton wave radiant
point. At this time, the perturbation profiles of the Moreton wave
are steepening (cf.\ Fig.~\ref{img:rb9-11rb4-6}). The surface of
these extended source sites is located at a distance $d\approx
60-200$~Mm from the Moreton wave radiant points. The first value is
determined from the type~II occurrence, while the second value is
given by the first Moreton wavefront appearance in this direction.

A strong relation between the CME lateral expansion and the
occurrence of coronal shock waves was already suggested by radio
observations. \cite{pohjolainen01} reported about a halo CME event,
accompanied by a Moreton and an EIT wave, in which the appearance of
new radio sources were attributed to the arc expansion of the CME
magnetic field. Finally, the encounter of the propagating wave with
magnetic field structures caused the opening of field lines.
\cite{maia99} related the systematic occurrence of radio sources
during the early evolution of a CME to the successive
destabilization or interaction of loop systems which resulted in
large-scale coronal restructering and finally CME formation. Similar
conclusions are drawn by \cite{temmer09} comparing the Moreton wave
observations of January, 17 2005 and the related type II burst
resulting in an outcoming analytical MHD model.

\subsection{Coronal shock characteristics}

It can be presumed that at low heights the coronal shock is a
quasi-perpendicular fast-mode MHD shock \citep{Mann:1999}.
Considering jump relations at the perpendicular shock front
\citep[e.g.,][]{Priest:1982}, the shock magnetosonic Mach number
$M_{ms}=v_{c}/v_{ms}$ (where $v_{c}$ is the coronal shock velocity
and $v_{ms}$ the magnetosonic speed) can be expressed as
\begin{equation}
 M_{ms}= \sqrt{\frac{X_c (X_c +5+5\beta_c)}{(4-X_c)(2+5\beta_c/3)}}\,.
\end{equation}
where $\beta_c$ is the ambient coronal plasma-to-magnetic pressure
ratio and $X_c=\rho_{c2}/\rho_{c1}$ the density jump at the shock
front. For the specific-heat ratio (the polytropic index) we have
taken $\gamma=5/3$. Using $X_c=1.32$\,--\,1.56 for the density jump,
as derived from the type II band-split observations (cf.\
Sect.~3.6), we find $M_{ms}=1.23$\,--\,1.42 for $\beta_c=1$, and
$M_{ms}=1.25$\,--\,1.45 for $\beta_c=0.01$.

The pressure jump at the shock front, $P_c=p_{2c}/p_{1c}$, can be
written as
\begin{equation}
 P_c = ~ 1 + \frac{5M_{c}^2(X_c-1)}{3X_c} - \frac{X_c^2-1}{\beta_c}\,,
\end{equation}
where $M_c$ represents the sound Mach number of the coronal shock,
which for $\gamma=5/3$ reads:
\begin{equation}
 M_c=~M_{ms}~\sqrt{1+\frac{6}{5\beta_c}}\,.
\end{equation}
Using $X_c=1.32$\,--\,1.56, we find $P_c=2.82$\,--\,9.34 for
$\beta_c=0.01$, and $P_c=1.61$\,--\,2.22 for $\beta_c=1$. Thus, the
type II burst observations indicate that the pressure jump ranges
between $\approx$\,2 and 10. The pressure jump at the shock front
causes an abrupt pressure increase at the top of a given
chromospheric element after the shock sweeps over it. Consequently,
also a pressure discontinuity forms between the perturbed corona and
the yet undisturbed chromosphere. The appearance of the pressure
discontinuity at the top of the chromosphere implies that a
downward-traveling shock should be formed. Behind the shock front
the plasma is compressed and moves downwards (including the
corona/chromosphere ``interface''). As the shock penetrates deeper,
a successively larger portion of the chromosphere is compressed and
set into motion, increasing the H$\alpha$ blue-minus-red Doppler
signal at the leading part of the Moreton wave.

\cite{vourlidas03} obtained from a CME case study that the
speed and density of the CME front and flanks, derived from
coronagraphic white light images, were consistent with the existence
of a shock. Comparison of the observations with MHD simulations
confirmed that the observed wave feature is a density enhancement
from a fast-mode MHD shock (for further similar white light
observations of coronal shocks see also \cite{ontiveros09}).
\cite{yan06} found from radio imaging observations that the emitting
sources of type~III bursts were located at the compression region
between the CME flanks and the neighboring open field lines. The
compression region was revealed as a narrow feature in white light,
and was interpreted as a coronal shock driven by the CME lateral
expansion. It is occasionally also observed as a type~IV-like burst
propagating colaterally with the Moreton wave
\citep{Vrsnak:2005,dauphin06}. Especially important are observations
of broadening and intensity changes of UV emission lines in front of
CMEs, associated with type II bursts, since such measurements
provide an insight into the physical state of plasma in the
compression region \citep{raymond00, mancuso02, ciaravella05}.
Occasionally, the plasma parameters can be inferred also from the
soft X-ray observations \citep{narukage02}.

\section{Summary and Conclusions}

We analyzed the intense Morton wave of October 28, 2003 together
with its associated phenomena: the X17.2/4B flare event, the fast
halo CME ($v\approx$~2500~km~s$^{-1}$), the EIT wave, the radio type
II burst, and coronal dimmings, in order to study the wave
characteristics and its driver. The main findings of the study and
our interpretation are summarized in the following:

\begin{enumerate}
  \item The Moreton wave propagates in almost all directions over the whole solar disk, and can be observed for a period as long
  as 11~min (during 11:02--11:13~UT). These are both unusual properties of a Moreton
  wave. But we note that a wave with similar characteristics (global propagation,
  $v\approx$~1100$-$1200~km~s$^{-1}$) was launched from the same AR on
  2003 October 29 in assocation with an X10 flare/CME event
  \citep[][]{Balasubramaniam:2007}. We studied the wave kinematics in detail for four different propagation directions finding
  mean velocities in the range 900--1100~km~s$^{-1}$. Typical values  of the sound speed and the
  Alfv$\acute{\hbox{e}}$n speed in the solar corona are
  180~km~s$^{-1}$ and 300--500~km~s$^{-1}$ (Mann et al.\ 1999), respectively, which
  corresponds to a magnetosonic Mach number of $\approx$2--3 of
  the Moreton wave under study, if interpreted as a propagating fast-mode
  shock. The band-split of the associated type II burst was determined to
  0.15--0.25, from which we inferred magnetosonic Mach number
  $M_s=1.23-1.45$.
  \item We find two radiant points (source centers) for the Moreton wave on
  opposite East-West edges of the source active region (AR~10486). This has not been
  reported before and indicates that either two independent waves were launched simultaneously, or that the wave
  was caused by the expansion of an extended source with a preferred axis.
  \item The associated coronal wave (``EIT wave'') was observed by the EIT
  instrument in one frame. The determined EIT wavefront lies on the extrapolated kinematical curve of the
  Moreton wave, identifying both the EIT wave and the H$\alpha$ Moreton
  wave as different signatures of the same disturbance, according to Uchida's theory.
  \item In two propagation directions, a deceleration of $-2960$~m~s$^{-2}$ and $-1420$~m~s$^{-2}$
  was observed in the distance-time diagrams of the H$\alpha$ Moreton wavefronts.
  In the initial phase, the perturbation profiles show amplitude
  growth and steepening of the disturbance. Later on, they display a
  broadening, amplitude decrease, and signatures of chromospheric relaxation appear at the trailing segment of the
  disturbance. This indicates that the coronal perturbation is
  a freely propagating large-amplitude simple-wave. The
  width of the disturbance is in the range of 50--150~Mm.
  \item The extrapolated start time of the wave fits roughly with the extrapolated
  start time of the CME observed in LASCO (but we note that the uncertainty of the CME start time is of the order of 10~min)
  but occurs slightly before (2~min) the first HXR peak of the flare impulsive phase (cf.\ Fig.~\ref{img:plot_HXR}).
  \item The associated bipolar coronal dimmings, which are generally interpreted as footprints of the expanding CME and associated coronal mass depletion,
  appear on the same opposite East-West edges of AR~10486 as the Moreton wave ignition centers do. The bipolar dimmings show a fast increase
  during the impulsive phase, followed by a much slower evolution, which indicates a characteristic restricted growth.
  The expansion velocity of the dimming region is $\approx$100~km~s$^{-1}$ at the time, where the Moreton wave is launched.
  \item The position of the type II burst source is co-spatial with
  the Moreton wavefront in direction 4$-$6. The back-extrapolated start time for the type II burst
  is $\approx$11:02:30~UT, i.e., somewhat later than the start time derived for the Moreton wave with respect to its radiant point.
  This suggests that the perturbation was not launched from a point-like source but from
  an extended source whose surface was at a distance $\gtrsim$60~Mm from the back-extrapolated radiant point.
  The launch time of the type II burst, which indicates the shock front formation, and the Moreton wave
  perturbation profile steepening coincide within the measurement errors.
\end{enumerate}

From several case studies
\citep[e.g.][]{vourlidas03,Mancuso:2003,yan06,Cho:2008,ontiveros09}
it was derived that radio as well as white light signatures in
coronagraphs provide evidence that the lateral expansion of a CME is
able to produce shocks in the corona. The results of the study,
presented in this paper, support a scenario in which the launch of
the observed Moreton and EIT wave is initiated by the lateral
expansion of the CME. In such a case we expect that the perturbation
is driven only over a short time/distance, as the CME
lateral-expansion range is limited \citep[for the theoretical
background see, e.g.][]{zic08,temmer09}. Thus, the disturbance would
evolve into a freely propagating wave, which is consistent with the
observed perturbation characteristics. The behavior of CME expansion
(radial and lateral direction) at low coronal heights might be an
important condition for the formation of coronal shock waves. 

\begin{acknowledgements}
We thank Monique Pick and Jean-Marie Malherbe for providing the
H$\alpha$ data from the Meudon Observatory. We also thank the SoHO,
LASCO, TRACE, INTEGRAL, AIP and Nan{\c c}ay teams for their open
data policy. N.M. acknowledges support by the Austrian Research
Program MOEL-PLUS. M.T. acknowledges project APART~11262 of the
Austrian Academy of Sciences. A.V. acknowledges the Austrian Fonds
zur F\"orderung der wissenschaftlichen Forschung under project FWF
P20867-N16.
 \end{acknowledgements}

\bibliographystyle{aa}

\end{document}